\documentclass[a4paper,11pt]{emulateapj}
\usepackage{amssymb, latexsym}
\usepackage{graphicx}
\usepackage{amsmath} 
\usepackage{longtable}

\def\meanz{\langle z \rangle}
\def\sigz{\sigma_{z}}

\usepackage{footmisc}
\usepackage{perpage} 
\MakePerPage{footnote} 

\begin{document}
\title{Mapping the Galaxy Color-Redshift Relation: Optimal Photometric
  Redshift Calibration Strategies for Cosmology Surveys}
\author{Daniel Masters\altaffilmark{1}, Peter Capak\altaffilmark{2},
  Daniel Stern\altaffilmark{3}, Olivier Ilbert\altaffilmark{4}, Mara Salvato\altaffilmark{5},
  Samuel Schmidt\altaffilmark{6}, Giuseppe Longo\altaffilmark{7}, Jason Rhodes\altaffilmark{3,13}, 
  Stephane Paltani\altaffilmark{8}, Bahram Mobasher\altaffilmark{9}, Henk Hoekstra\altaffilmark{10},
  Hendrik Hildebrandt\altaffilmark{11}, 
  Jean Coupon\altaffilmark{8}, Charles Steinhardt\altaffilmark{1}
  Josh Speagle\altaffilmark{12}, 
  Andreas Faisst\altaffilmark{1}, Adam Kalinich\altaffilmark{14}, Mark Brodwin\altaffilmark{15}
  Massimo Brescia\altaffilmark{16}, Stefano Cavuoti\altaffilmark{16}}

\altaffiltext{1}{Infrared Processing and Analysis Center, California
  Institute of Technology,
  Pasadena, CA 91125}
\altaffiltext{2}{Spitzer Science Center, California Institute of
  Technology, Pasadena, CA 91125, USA}
\altaffiltext{3}{Jet Propulsion Laboratory, California Institute of
  Technology, Pasadena, CA 91109}
\altaffiltext{4}{Aix Marseille Universit´e, CNRS, LAM (Laboratoire d’Astrophysique de Marseille) UMR 7326, 13388, Marseille,
France}
\altaffiltext{5}{Max-Planck-Institut für extraterrestrische Physik,
  Giessenbachstrasse, D-85748 Garching, Germany}
\altaffiltext{6}{Department of Physics, University of California,
  Davis, CA 95616}
\altaffiltext{7}{Department of Physics, University Federico II, via
  Cinthia 6, 80126 Napoli, Italy}
\altaffiltext{8}{Department of Astronomy, University of Geneva
  ch. d’cogia 16, CH-1290 Versoix}
\altaffiltext{9}{Department of Physics and Astronomy, University of California, Riverside, CA 92521}
\altaffiltext{10}{Leiden Observatory, Leiden University, PO Box 9513, 2300 RA, Leiden, the Netherlands}
\altaffiltext{11}{Argelander-Institut f{\"u}r Astronomie, Universit{\"a}t Bonn, Auf dem H{\''u}gel 71, 53121 Bonn, Germany}
\altaffiltext{12}{Department of Astronomy, Harvard University, 60 Garden Street, MS 46, Cambridge, MA 02138, USA}
\altaffiltext{13}{Kavli Institute for the Physics and Mathematics of the Universe The University of Tokyo, Chiba 277-8582, Japan}
\altaffiltext{14}{Massachusetts Institute of Technology, Cambridge, MA
  02139}
\altaffiltext{15}{Department of Physics and Astronomy, University of Missouri, Kansas City, MO 64110, USA}
\altaffiltext{16}{Astronomical Observatory of Capodimonte - INAF, via Moiariello 16, I-80131, Napoli, Italy}

\begin{abstract}

Calibrating the photometric redshifts of $\gtrsim10^{9}$
galaxies for
upcoming weak lensing cosmology experiments is a major
challenge for the astrophysics community. The path to obtaining the required spectroscopic
redshifts for training and calibration is daunting, given the anticipated depths of the
surveys and the difficulty in obtaining secure redshifts
for some faint galaxy populations. Here we present an analysis of the
problem based on the \emph{self-organizing map}, a method of mapping
the distribution of data in a high-dimensional space and projecting it
onto a lower-dimensional representation. We apply this method
to existing photometric data from the COSMOS survey selected to approximate the anticipated \emph{Euclid} weak
lensing sample, enabling us to robustly map the empirical distribution of galaxies in the
multidimensional color space defined by the expected \emph{Euclid}
filters. Mapping this multicolor distribution lets us determine where -- in
galaxy color space -- redshifts from current 
spectroscopic surveys exist and where they are
systematically missing. Crucially, the method lets us determine
whether a spectroscopic training sample is representative of the full
photometric space occupied by the galaxies in a survey. We explore optimal sampling techniques and estimate the additional spectroscopy needed
to map out the color-redshift relation, finding
that sampling the galaxy distribution in color space in a systematic
way can efficiently meet the calibration requirements. While the analysis
presented here focuses on the \emph{Euclid} survey, similar analysis
can be applied to
other surveys facing the same calibration challenge, such as DES,
LSST, and \emph{WFIRST}. 

\end{abstract}
\maketitle

\section{Introduction}

Upcoming large-scale surveys such as LSST, \emph{Euclid} and \emph{WFIRST} will
measure the
three-dimensional cosmological weak lensing shear field from
broadband imaging of billions of galaxies. Weak lensing is widely
considered to be one of the most promising probes of the growth of dark
matter structure, as
it is sensitive to gravitation alone and requires
minimal assumptions about the coupling of dark matter and 
baryons \citep{Bartelmann01,Weinberg13}. Moreover, weak lensing tomography is sensitive to the dark
energy equation of state through its impact on the growth of structure with time
\citep{Hu99}. However, it is observationally demanding: in addition to requiring accurately measured shapes 
for the weak lensing sample, robust 
redshift estimates
to the galaxies are needed in order to reconstruct the three-dimensional
matter distribution. Because it is infeasible to obtain spectroscopic
redshifts (spec-z's) for the huge numbers of faint galaxies
these studies will detect, photometric
redshift (photo-z) estimates derived from imaging in some number of broad filters will be required for nearly all
galaxies in the weak lensing samples. 

Photo-z
estimation has become an indispensable tool in
extragalactic astronomy, as the pace of galaxy detection in imaging
surveys far outstrips the rate at which follow-up
spectroscopy can be performed. While photo-z techniques have grown
in
sophistication in
recent years, the requirements for
 cosmology present novel challenges. In particular,
cosmological parameters derived from weak lensing are sensitive to
small, systematic errors in the photo-z estimates \citep{Ma06,
  Huterer06}. Such biases are generally much smaller than
the random scatter in photo-z estimates \citep{Dahlen13}, and are of
little consequence for galaxy evolution studies; however, they
can easily dominate all other uncertainties in weak lensing
experiments \citep{Newman15}. In addition to weak lensing cosmology, accurate and well-characterized
  photo-z's will be crucial to other cosmological
  experiments. For example, baryon acoustic oscillation
  (BAO) experiments that rely on redshifts measured from faint
  near-infrared grism spectra will often have to resort to photo-z's
  in order to determine the correct redshift assignment for galaxies with only a single detected
  line. Well-characterized photo-z estimates will be needed to
  correctly account for any errors
  thus introduced.

There are two key requirements placed on the
photo-z estimates for weak lensing cosmology. First, redshift
estimates for individual objects must have sufficient precision to correct for
intrinsic galaxy shape alignments as well as other
potential systematics arising from physically associated galaxies
that may affect the interpretation
of the shear signal. While not
trivial, meeting the
requirement on the precision of individual photo-z
estimates ($\sigz < 0.05(1+z)$ for \emph{Euclid},
\citealp{Laureijs11}) should be achievable \citep{Hildebrandt10}. The
second, more difficult, requirement is that the overall redshift
distributions $N(z)$ of galaxies in $\sim$10--20 tomographic bins
used for the shear analysis
must be known with high accuracy. Specifically, the mean redshift $\meanz$ of the $N(z)$ distribution must be constrained to
better than $2\times10^{-3}(1+z)$ in order to interpret the amplitude
of the lensing signal and achieve acceptable error levels on the
cosmological parameter estimates \citep{Huterer06, Amara07, Laureijs11}. Small biases in the photo-z
estimates, or a relatively small number of objects with
catastrophically incorrect photo-z's, can cause
unacceptably large errors in the estimated $N(z)$
distribution. Photo-z estimates alone are not sufficient to meet
this requirement, and spectroscopic calibration samples will be
needed to ensure low bias in the $N(z)$ estimates. The significant
difficulties associated with this requirement are summarized by \citet{Newman15}.

The most straightforward approach to constrain $N(z)$ is to measure it
directly by random spectroscopic sampling of galaxies in each
tomographic redshift
bin \citep{Abdalla08}. The total number of spectra needed to meet the requirement is
then set by the central limit theorem. For upcoming ``Stage IV''
cosmology surveys (LSST, \emph{Euclid}\footnote[2]{http://www.euclid-ec.org}, and \emph{WFIRST}) it is estimated
that direct measurement of $N(z)$ for the
tomographic bins would require total spectroscopic samples of
$\sim$30,000--100,000 galaxies, fully representative in flux,
color, and spatial distribution of the galaxies used to measure the
weak lensing shear field (e.g., \citealp{Ma08}, \citealp{Hearin12}). Moreover, the spectroscopic redshifts
would need to have a very high success rate ($\gtrsim$99.5\%), with no subpopulation of
galaxies systematically missed in the redshift survey. \citet{Newman15} note that current
deep redshift surveys fail to obtain secure redshifts
for $\sim$30--60\% of the targeted galaxies; given the depths of the planned
dark
energy surveys, this ``direct'' method of calibrating the redshifts
seems to be unfeasible. 

Because of the difficulty of direct spectroscopic calibration,
\citet{Newman15} argue that the most realistic method of meeting
the requirements on $N(z)$ for the dark energy experiments may be some
form of  
spatial 
cross-correlation of photometric samples with a reference spectroscopic
sample, with the idea that the power in the
cross-correlation will be highest
when the samples match in redshift 
\citep{Newman08, Schmidt13, Rahman15}. This approach shows significant promise, but
is not without
uncertainties and potential systematics. For example, it requires
assumptions regarding the growth of structure and galaxy bias with
redshift, which may be covariant with the cosmological inferences
drawn from the weak lensing analysis itself. Further work may clarify
these issues and show that the technique is indeed viable for upcoming
cosmological surveys. However, it seems safe to say that this method cannot
\emph{solely} be relied on for the weak lensing missions, particularly as
at least two approaches will be needed: one to calibrate $N(z)$ for the
tomographic bins, and another to
test and validate the calibration. 

In light of these arguments, it is clear that targeted spectroscopic training and calibration samples will
have to be obtained to achieve the accuracy in the $\meanz$
estimates of tomographic bins required by the
weak lensing missions. Moreover, careful optimization of these
efforts will be required to
make the problem tractable. Here we present a technique, based on the simple but powerful \emph{self-organizing map}
\citep{Kohonen82, Kohonen90}, to map the empirical distribution of
galaxies in the multidimensional color space defined by a 
photometric survey. Importantly, this technique provides us with a
completely data-driven understanding of what constitutes a representative
photometric galaxy sample. We can thereby evaluate whether a
spectroscopic sample used for training and calibration spans
the full photometric parameter space; if it does not, there will be
regions where
the photo-z results are untested and untrained. Machine learning--based
photo-z algorithms, in particular, depend critically on representative
spectroscopic training
sets, and their performance will be degraded in regions of color space
without spectroscopic coverage\footnote[3]{This dependence on training
  sample representativeness tends to be obscured by the
photo-z versus spec-z plots most often used to illustrate the quality of
photo-z algorithms, which (of necessity) only show results for the
subset of
galaxies with known spectroscopic redshifts. Of course, those are also
the
galaxies for which similar training objects exist.}
\citep{Collister04, Hoyle15}. 

We show that the empirical color
mapping described here can be used to optimize
the training and calibration effort by focusing
spectroscopic effort on regions of galaxy parameter space
that are currently poorly explored, as well as
regions with a less certain mapping to redshift. Alternatively, we can use
the technique to identify and discard specific regions of color space for which
spectroscopy will prove to be too
expensive, or for which the redshift uncertainty is too large. In
effect, the method lets us systematize our
understanding of the mapping from color to redshift. By doing so, the number of
spectroscopic redshifts needed to calibrate $N(z)$ for the weak
lensing tomographic bins can be
minimized. This approach will also naturally produce a ``gold standard'' training
sample for machine learning algorithms.

The technique we adopt also provides insight into the nature
of catastrophic photo-z failures by illustrating regions of color
space in which the mapping between color and redshift becomes
degenerate. This is possible because the self-organized map is
topological, with
nearby regions representing similar objects, and widely separated
regions representing dissimilar ones. In addition, template-fitting
photo-z codes can potentially be refined with the map, particularly
through the development of data-based 
priors and by using the empirical color mapping to test and refine the
galaxy template sets
used for fitting.

Here our focus is on the \emph{Euclid} survey, one of the three
Stage IV dark energy surveys planned for the next decade,
the other two being LSST and \emph{WFIRST}. \emph{Euclid}
  will consist of a 1.2~meter space telescope operating at L2, which
  will be used to 
  measure accurate shapes of galaxies out to $z$$\sim$2 over
  $\sim$15,000 deg$^2$ with a single, broad (\emph{riz}) filter. These
  observations will reach an AB magnitude of~$\simeq$24.5 (10$\sigma$). In
  addition to these observations, a near-infrared
  camera on \emph{Euclid} will obtain \emph{Y}, \emph{J}, and \emph{H}
  band photometry to AB magnitude $\simeq$24~(5$\sigma$), which,
  together with complementary ground-based optical data, will be used for
  photo-z determination. The mission will also constrain cosmological
  parameters using BAO and redshift space distortions (RSD), using
  redshifts obtained with a low-resolution
  grism on the near-infrared camera. A more detailed description of
  the survey can be found in \citet{Laureijs11}.

For this work, we assume that \emph{Euclid} will 
obtain $ugrizYJH$ photometry for photo-z estimation. We select galaxies from the
COSMOS survey \citep{Scoville07} that closely approximate the \emph{Euclid}
weak lensing sample, with photometry in similar bands and at similar
depths as the
planned \emph{Euclid} survey. While our focus is on \emph{Euclid}, the method
we present is general
and directly
applicable to other weak lensing surveys facing the same calibration
problem.

This paper is organized as follows. In \S2 we give an overview of
the methodology used to map the galaxy multicolor space. In \S3 we
discuss the galaxy sample from the COSMOS survey used to approximate
the anticipated \emph{Euclid} weak lensing sample. In \S4 we describe the
self-organizing map algorithm and its implementation for this application. In \S5 we discuss the map in detail, including what it reveals
about the current extent of spectroscopic coverage in galaxy multicolor space. In \S6 we address
the problem of determining the spectroscopic sample needed to meet the
weak lensing requirement, and in \S7 we conclude with a discussion.

\section{Overview: Quantifying the Empirical Distribution of Galaxies in Color Space}

Galaxies with imaging in a set of $N$ filters will
follow some distribution in the multidimensional space (of dimension $N-1$) defined by the
unique colors measured by the filters. These colors together determine the
shape of the low-resolution spectral energy distribution (SED) measured by the filters.
Henceforth, we will call
the position a galaxy occupies in color space simply its color, or
$\vec{C}$. For example, the \emph{Euclid}
survey is expected to have eight bands of photometry ($ugrizYJH$)\footnote[4]{This
  will be the case in the region overlapping with the LSST
  survey. We note that \emph{Euclid} will also have a broad ($riz$) filter
  that will be used for
  the weak lensing shape measurements; our assumption here is that it will
  not add significant value to the photo-z estimates.},
and therefore a galaxy's  position in color space is uniquely determined by seven
colors: $u-g, g-r, ..., J-H$. Galaxy color
is the primary driver of photometric redshift estimates:
template-based methods predict $\vec{C}$ for different
template/redshift/reddening combinations and assign redshifts to galaxies based on where the models
best fit the observed photometry, while machine learning methods
assume the existence of a mapping from $\vec{C}$ to
redshift, and attempt to discover it using spectroscopic training
samples. 

Our goal here is to empirically map
the distribution of galaxies in the color space
defined by the anticipated \emph{Euclid} broadband filters. We  
refer to this distribution as $\rho{(\vec{C})}$. Once we understand how galaxies are distributed in color
space, optimal methods of sampling the distribution with spectroscopy can be
developed to make an informed calibration of the color-redshift
relation.

The general problem of mapping a high-dimensional data distribution arises in
many fields. Because the volume of
the data space grows exponentially with the number of
dimensions, data rapidly becomes sparse as the dimensionality
increases. This effect -- the so-called ``curse of dimensionality''
\citep{Bellman57} --
makes normal data sorting strategies impractical. A number of algorithms,
collectively referred to as nonlinear dimensionality reduction (NLDR),  have been developed to address
this problem by projecting high-dimensional
data onto a lower-dimensional representation, thus
facilitating visualization and analysis of relationships that exist in
the data. 

We
adopt the self-organizing map algorithm, described in more
detail in \S4. As emphasized by \citet{Geach12}, self-organized mapping is a
powerful, empirical method to understand the multidimensional 
distributions common in modern astronomical surveys.  Two primary
motivations for choosing this technique over others
are the relative simplicity of the algorithm and the highly visual
nature of the resulting map, which facilitates human understanding of the data.

\section{Approximating the \emph{Euclid} Weak Lensing Sample with COSMOS Data}

We use multiwaveband data from the COSMOS survey \citep{Capak07} to provide a close
approximation to the expected \emph{Euclid} weak lensing data. Photo-z
estimates for the \emph{Euclid} sample will rely on three
near-infrared filters on the telescope ($YJH$), reaching an AB depth of
24 mag (5$\sigma$) for point sources, as well as complementary ground-based imaging in the optical, which we
assume will consist of $ugriz$ imaging with LSST (in the northern sky the ground-based
imaging data may be restricted to $griz$, affecting the analysis
somewhat but not changing the overall conclusions).

To provide a close analog to the expected \emph{Euclid} data, we use COSMOS
$u$
band imaging from CFHT, $griz$ imaging from Subaru Suprime Cam, and
$YJH$ imaging from the UltraVista survey \citep{McCracken12}, spanning a 1.44~deg$^{2}$
patch of COSMOS with highly uniform depth. We apply a flux cut to the
average flux measured across the Subaru $r$, $i$ and $z$ bands to match the
expected depth limit of the single, broad visible filter \emph{Euclid} will
use for the weak lensing shear measurement. The resulting ``Euclid
analog'' sample
consists of
131,609 
objects from COSMOS.

\begin{figure*}[htp]
        \centering
	\includegraphics[width=\linewidth]{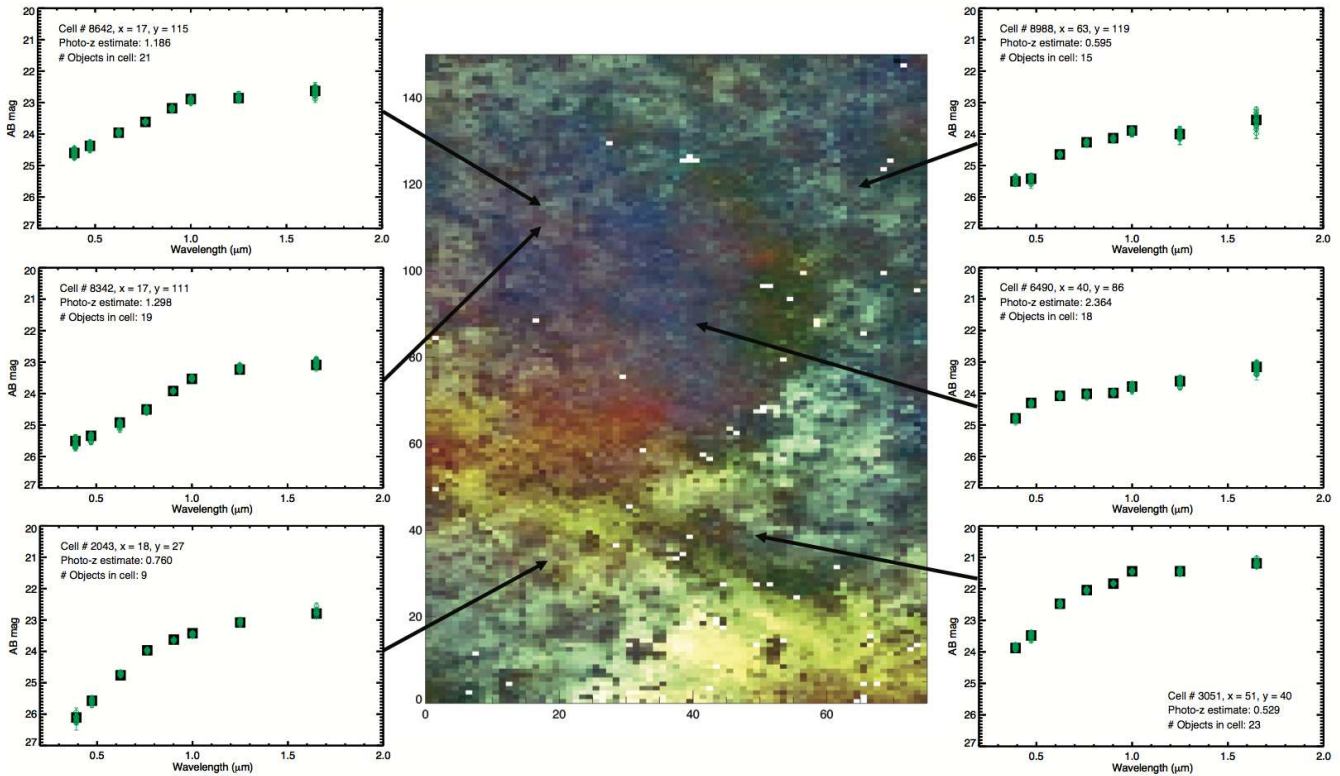}
	\caption{The 7-color self-organized map (SOM) generated from
          $\sim$131k galaxies from the COSMOS survey, selected to be representative of the anticipated
          \emph{Euclid} weak lensing sample. In the center
        is the $75\times150$ map itself, which encodes
        the empirical \emph{ugrizYJH} spectral energy distributions (SEDs) that appear in the data. The map is colored here by converting the \emph{H},
       \emph{i}, and \emph{u} band photometry of the cells
       to analogous RGB values, while the brightness is scaled to reflect the average
        brightness of galaxies in different regions of color space.  On the sides we show examples of
       8-band galaxy SEDs represented by particular cells, whose
       positions in the map are 
      indicated with arrows. The cell SEDs are shown as black
      squares. The actual SEDs (shifted to line up
      in $i$-band magnitude) of galaxies associated with the cells are
      overlaid as green diamonds. Between 9 and 23 separate galaxy SEDs are
      plotted for each of the cells shown, but they are similar enough
      that they are hard
      to differentiate on this figure. A key feature of the map is that it is
      topological, in the sense that nearby cells represent
      objects with similar SEDs, as can be seen from the two example
      cells shown in the upper left. Note that the axes of the SOM do not correspond
      to any physical quantity, but merely denote positions of cells
      within the map and are shown to ease comparison between figures.}
\label{figure:som}
\end{figure*}

\section{Mapping Galaxy Color Space with the Self-Organizing Map}

The self-organizing map (SOM, \citealp{Kohonen82, Kohonen90}) is a neural network model
widely used to map and identify correlations in high-dimensional
data. Its use for some astronomical applications has been explored
previously (see, e.g., \citealp{Naim97, Brett04, Way12, Fustes13, Kind14}). The algorithm uses unsupervised, competitive learning of
``neurons'' to project high-dimensional data onto a lower-dimensional grid. The SOM algorithm can be thought of as a type
of nonlinear principal component analysis, and is also similar in
some respects to the k-means clustering algorithm \citep{MacQueen67}. In contrast to
these and other methods, the SOM preserves the topology of the high-dimensional data in
the low-dimension representation. Similar objects are thus grouped
together on the self-organized map, and clusters that exist in the
high-dimensional data space are reflected in the lower-dimensional
representation.  This
feature makes the maps visually understandable and thus useful for 
identifying
correlations that exist in high-dimensional data. More detailed
descriptions of the algorithm and its variants can be found in a
number of references (see, e.g., \citealp{Vesanto02, Kind14}).

\begin{figure*}[htb]
\centering
    \includegraphics[width=\linewidth]{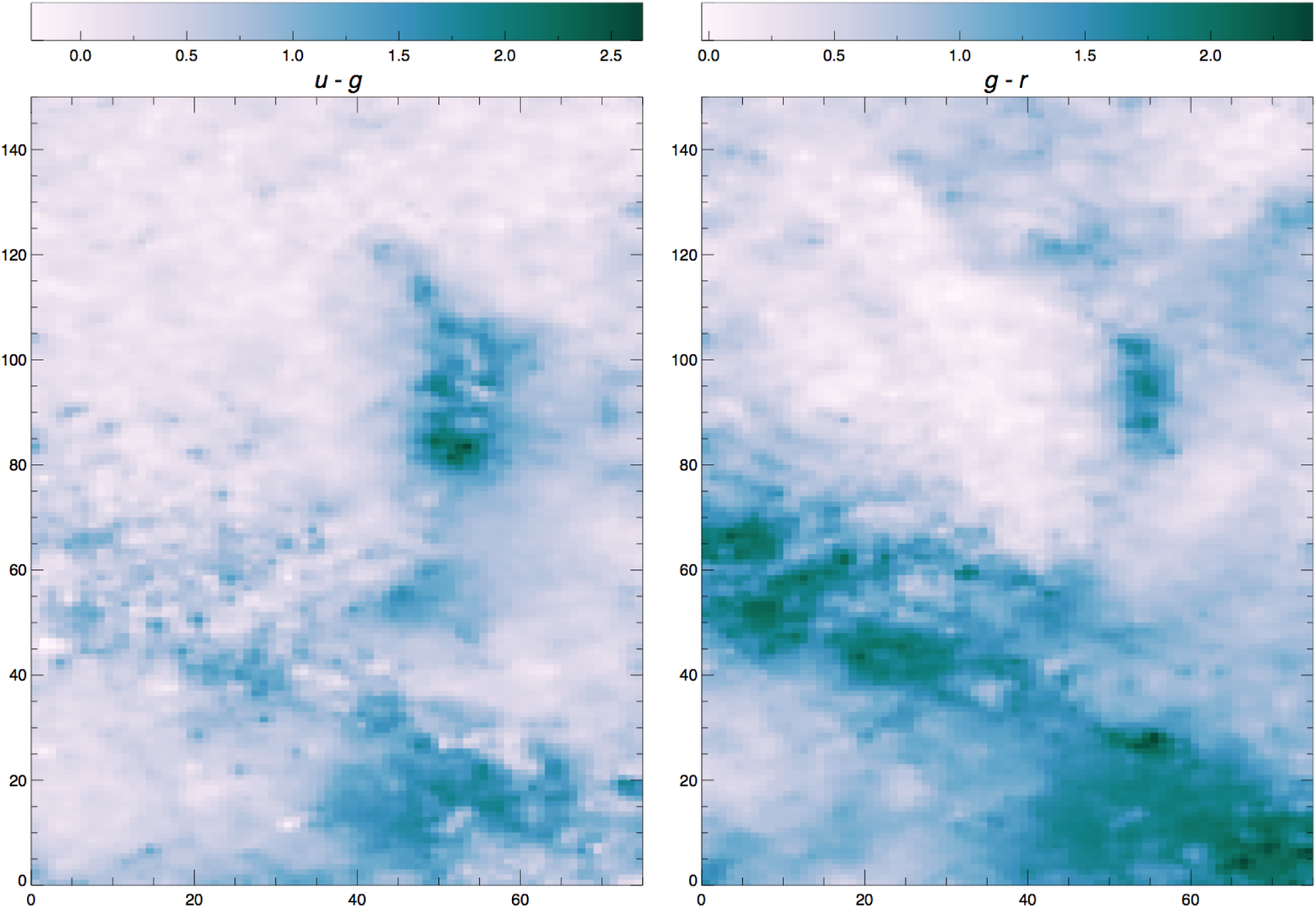} 
  \label{figure:colors}
  \caption{The variation of two colors along the
          self-organizing map: $u-g$ on the left and $g-r$
          on the right. In the language of machine learning, these are
        ``features'' in the data that drive the overall structure of
        the map. The well-known Lyman break is evident for galaxies at
      $2.5\lesssim z \lesssim 3$ in $u-g$ and $3 \lesssim z \lesssim 4$ in $g-r$ (around x=50, y=90). The regions
      with red $g-r$ color spreading diagonally across the lower part
      of the map are a combination of passive galaxies and dusty
      galaxies at lower redshift.}
\end{figure*}

The SOM consists of a fixed number of cells
arranged on a grid. The grid can be of arbitrary
dimension, although two-dimensional grids are most common as they are
the easiest to visualize. Each cell in the grid is assigned a 
weight vector $\vec{w}$ having the same number of dimensions as the
training data. This vector can be thought of as pointing to a particular
region of the multidimensional parameter space occupied by the
data. The weight vectors are initialized prior to training,
either randomly or by sampling from the input data. The training of the map is unsupervised, in the sense that the 
output variable of interest (here, redshift) is not
considered. Only the input attributes (galaxy
photometry) drive the training. We note that any measured galaxy property (size, magnitude,
shape, environment, surface brightness, etc.) could be used in the training. We consider only
colors here, as
these are the primary drivers of the photo-z estimates, and the
quantities most
physically tied to redshift. The other
properties mentioned can still
be used after the map has been created to identify and help
break redshift degeneracies within
particular regions of galaxy color space. 

Training proceeds by presenting the map with a random
galaxy from the training sample, which the cells ``compete'' for. The
cell whose weight vector most closely resembles the training galaxy is considered the
winner, and is called the Best Matching Unit, or
BMU. The BMU as well as cells in its neighborhood on the map are then modified to more
closely resemble the training galaxy.
This pattern is repeated for many training iterations, over which
the responsiveness of the map to new data gradually decreases, through
what is known as the learning rate function. Additionally, the extent
of the neighborhood around the BMU affected by new training data
shrinks with iteration number as well, through what is known as the neighborhood
function. These effects cause the map to settle to a stable
solution by the end of the training iterations. 

To compute the winning cell for a given training object, a distance metric must
be chosen. Most often, the Euclidean distance between the training
object $\vec{x}$ and the cell weight vector $\vec{w_{k}}$ is
used. With data of dimension $m$, this distance is given by: 
\begin{equation} d^{2}_{k} = d^{2}_{k}(\vec{x},\vec{w_{k}}) =
  \sum\limits_{i=1}^{m}(x_{i}-w_{k,i})^{2} \end{equation} 
However, dimensions with intrinsically larger error than others
will be overweighted in this distance metric. To account for
this, we instead use the reduced $\chi^{2}$
distance between the training object and the cell weight
vector. With $\sigma_{x_{i}}$ representing the uncertainty in the $i^{\mathrm{th}}$
component of $\vec{x}$, this becomes:
 \begin{equation} d^{2}_{k} = d^{2}_{k}(\vec{x},\vec{w_{k}}) =
\frac{1}{m}\sum\limits_{i=1}^{m}\frac{ (x_{i}-w_{k,i})^{2}}{\sigma_{x_{i}}^{2}} \end{equation}

The BMU is the cell minimizing
the $\chi^{2}$ distance. Once the BMU has been identified, the weight
vectors of cells in the map are updated with the
relation: \begin{equation}
  \vec{w_{k}}(t+1)=\vec{w_{k}}(t)+a(t)H_{b,k}(t)[\vec{x}(t)-\vec{w_{k}}(t)] \end{equation}
Here $t$ represents the current timestep in the training. The learning
rate function $a(t)$ is a monotonically decreasing function of the
timestep (with $a(t)\leq1$), such that the SOM becomes
progressively less responsive to new training data. With $N_{iter}$ representing the
total number of training iterations, we adopt the
following functional form for $a(t)$: \begin{equation} a(t) =
  0.5^{(t/N_{iter})} \end{equation}
The term $H_{b,k}$(t) is the value of the neighborhood function at the
current timestep for cell $k$, given that the current BMU is cell
$b$. This function is encoded as a normalized Gaussian kernel
centered on the BMU: \begin{equation} H_{b,k}(t) =
  e^{-D^{2}_{b,k}/\sigma^{2}(t)} \end{equation} Here $D_{b,k} $ is
the Euclidean distance on the map separating the $k^{\mathrm{th}}$ cell and the
current BMU. The
width of the Gaussian neighborhood function is set by $\sigma(t)$ and is given
by \begin{equation} \sigma(t) = \sigma_{s}
  (1/\sigma_{s})^{(t/N_{iter})} \end{equation} The starting value, $\sigma_{s}$,
   is large enough that the
  neighborhood function initially encompasses
most of the map. In practice, we set $\sigma_{s}$ equal to the 
the size (in pixels) of the smaller dimension of the rectangular map. The width of the neighborhood function shrinks by the end of training such that only the
BMU and cells directly adjacent to it are significantly affected by
new data. 

\subsection{Optimizing the map for the photo-z problem}

There is significant flexibility in choosing the parameters of the
SOM. Parameters that can be modified include the number
of cells, the topology of the map,
the number of training iterations, and the form and evolution of the learning rate and
neighborhood functions. Perhaps most influential is the number of
cells. The representative power of the map increases with
more cells; however, if too many
cells are used the map will 
overfit the data, modeling noise that does not reflect the 
true data distribution. Moreover, there is a significant computational cost to
increasing the number of cells. On the other hand, if too few cells are used,
individual cells will be forced to
represent larger volumes of color space, in which the mapping of color to
redshift is less well defined.

We explored a range of alternatives prior to
settling on the map shown throughout this work. A rectangular map
was chosen because this gives any principal component in the data a
preferred dimension along which to align. Our general guideline in
setting the number of cells was that the
map should have sufficient resolution such that the individual cells
map cleanly to redshift using standard photo-z codes. With 11,250
cells, the map bins galaxies into volumes, or ``voxels'', of color space of
comparable size as the photometric error on the data, with the result
that 
variations within each color cell generally do
not result in significant change in photo-z estimates. As we 
discuss in \S6, the true spread in galaxy redshifts within
each color cell is an important quantity to understand for the
calibration of $N(z)$.

\subsection{Algorithm implementation}
We implemented the SOM
algorithm in C for computational efficiency. The number of
computations required is sizable and scales with both 
the total number of cells and the number of training iterations. Optimizations are certainly
possible, and may be necessary if this algorithm is to be applied to
much larger photometric datasets. We initialized the values of the
cell weight vectors with random numbers drawn from a standard normal distribution. The number of training
iterations used was $2\times10^{6}$, as only minimal improvements in the
map were observed for larger numbers of iterations. At each iteration,
a random galaxy was
selected (with replacement) from the training sample to
update the map.

We applied the algorithm based on seven galaxy colors: $u-g$, $g-r$, $r-i$, $i-z$,
$z-Y$, $Y-J$, and $J-H$, which are analogous to the colors
that will be measured by \emph{Euclid} and used for photo-z estimation. The errors in the colors are computed
as the quadrature error of the photometric errors in the individual
bands. If a training object has a 
color that is not constrained due to bad photometry
in one or both of the relevant bands, we
ignore that color in the training iteration. Only the well-measured colors
  for that object are used both to find the BMU and update the
  corresponding colors of the cell
  weight vectors.  If a color represents an upper/lower limit, we penalize the
$\chi^{2}$ distance for cells
that violate the limit when computing the BMU, with a penalty that
varies depending on the size of the discrepancy between the limit and the cell color value.

\begin{figure}
        \centering
	\includegraphics[width=0.95\linewidth]{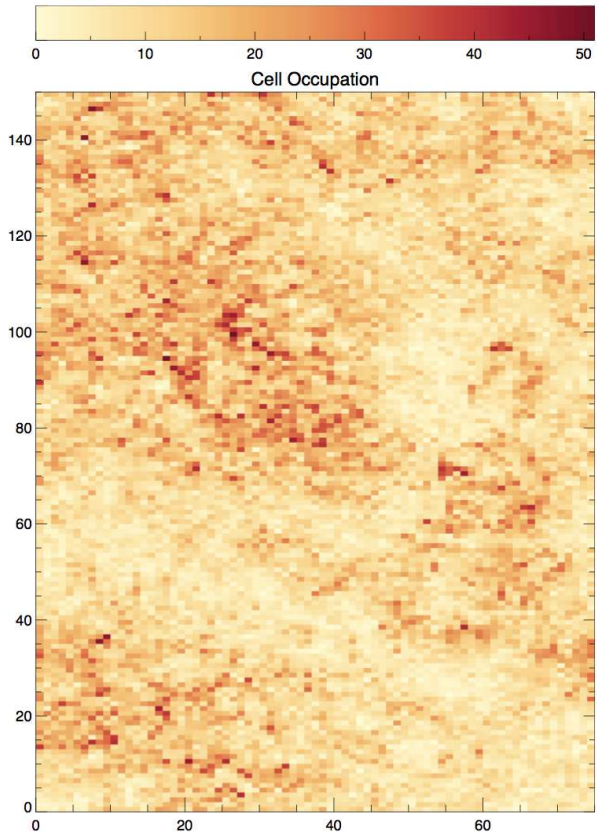}
	\caption{The SOM colored by the number of galaxies in the
          overall sample associating with each color cell. The coloration is effectively our
      estimate of $\rho{(\vec{C})}$, or the density of galaxies as a function
      of position in color space. \vskip 0.1cm}
\label{figure:occupation}
\end{figure}

\subsection{Assessing map quality}

Ideally, the SOM should be highly representative of the data,
in the sense that the SEDs of most galaxies in the sample are well-approximated by
some cell in the map. To assess the representativeness of the map we
calculate what is known as the average
quantization error over the entire training sample of $N$ objects: \begin{equation} \epsilon_{q} = 
  \frac{1}{N}\sum\limits_{i=1}^N ||\bold{x_{i}}-\bold{b_{i}}|| \end{equation} 

Here $\bold{b_{i}}$ is the best matching cell for the
\emph{i}$^{\mathrm{th}}$ training object. We find that the average
quantization error is 0.2 for the sample. The quantization error is the average vector distance between an object and
its best-matching cell in the map\footnote[5]{We do not use the
  $\chi^{2}$ distance for this test because of the discrete nature of
  the cells. Bright objects, or those with smaller photometric
  errors, will have artifically higher $\chi^{2}$ separation from
  their best-matching cell (even if the photometry matches well), making the metric less appropriate for
  assessing the representativeness of the map.}. Therefore, with seven colors used to generate 
the map, the average offset of a particular color (e.g.,
$g-r$) of a given galaxy from its corresponding cell in the map is $0.2/\sqrt{7}=0.08$~mag. 
Note that the map provides a straightforward way of identifying
unusual or anomalous sources. Such objects will be poorly represented
by the map due to their rarity -- in effect, they are unable to train
their properties into the SOM. Simply checking whether an
object is well represented by some cell in the map is therefore a way of testing whether
 it is ``normal'', and may be useful for flagging, for example, blended
 objects, contaminated photometry, or truly rare sources. 


\section{Analyzing the Color Space Map}

Figure \ref{figure:som} provides an overview of the SOM
generated from COSMOS galaxies, which encodes the 8-band SEDs that appear in the data with
non-negligible frequency. Note that the final structure of the map is
to some extent
random and depends on the initial conditions combined with the order in
which training objects are presented, but the overall
topological structure will be similar from run to run; this was
verified by generating and comparing a number of maps.\footnote[6]{See
  Appendix B for examples of alternate maps made with different initial
  conditions and training orders.} Figure~2 illustrates the variation of two colors ($u-g$ and
$g-r$) across the map, demonstrating how these features help drive the
overall structure. In the following analysis we probe the map by analyzing
the characteristics of the galaxies that associate best with each cell in
color space.

\begin{figure*}[htb]
\centering
    \includegraphics[width=\linewidth]{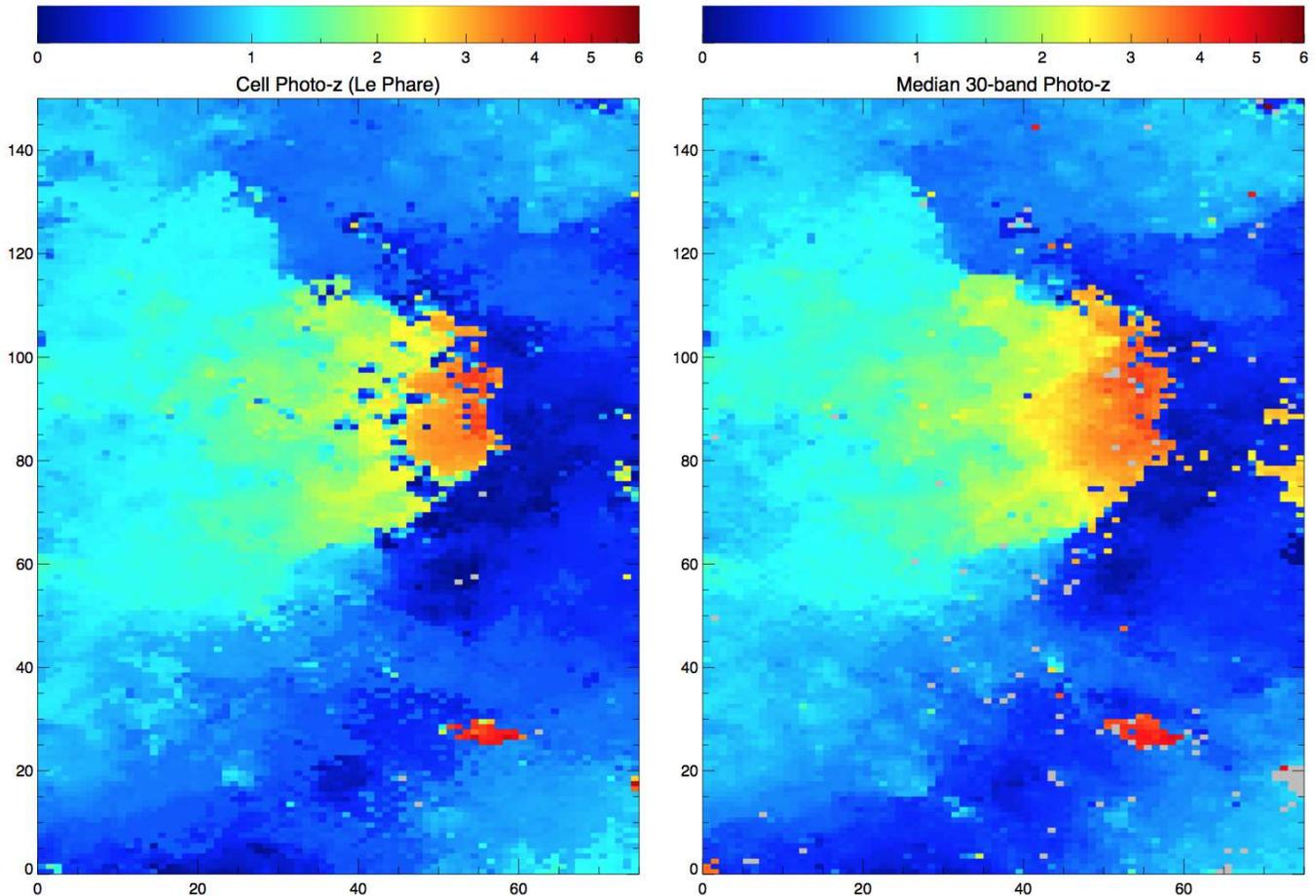}
  \label{figure:photz}
  \caption{Photo-z estimates across the map, computed in two
    ways. \emph{Left:} Photo-z's computed directly
          for each cell by applying the \emph{Le Phare} template fitting
          code to the 8-band photometry
          represented by the cells. \emph{Right:} Photometric redshifts for the
    cells computed as the median of the 30-band COSMOS photo-z's of the objects associated with each
          cell.}
\end{figure*}

\subsection{The distribution of galaxies in color space, $\rho{(\vec{C})}$}

In Figure \ref{figure:occupation} we show the self-organized map colored by
the number of galaxies associating best with each cell.  This coloration is
effectively our estimate of $\rho{(\vec{C})}$, the density of
galaxies as a function of position in color space. An important caveat
is that the density estimate derived from the
COSMOS survey data is likely to be affected to some degree by cosmic
variance (and perhaps, to a lesser extent, by shot noise). The
true $\rho{(\vec{C})}$ can ultimately be constrained firmly
with the wide-area survey data from LSST, 
\emph{Euclid}, and \emph{WFIRST}. However, 
the COSMOS-based $\rho{(\vec{C})}$ should be a close approximation of what the full
surveys will find.

\subsection{Photometric redshift estimates across the map}

Because the cells in the self-organizing map represent
galaxy SEDs that appear in the
data, we can compute photometric redshifts for them to
see how they are distributed in redshift. We used the \emph{Le
  Phare} template fitting code \citep{Arnouts99, Ilbert06} to compute
cell photo-z's. We used the cell weight vectors (converting the colors to
photometric magnitudes normalized in $i$-band) as inputs for \emph{Le
  Phare}, assigning realistic error bars to
these model SEDs based on the scatter in the photometry of
galaxies associated with each cell. The result of the photo-z fitting is shown on the left
side of Figure~4.

We also estimate redshifts on the map by computing the
median photo-z of the galaxies associated with each cell, using the
30-band photo-z estimates provided by the COSMOS survey \citep{Ilbert09}. These photo-z estimates take advantage of more
photometric information
than is contained in the eight \emph{Euclid}-like filters used to
generate the map. Nevertheless, as can be seen on the right side of Figure~4, the resulting map is quite smooth,
indicating that the eight \emph{Euclid} bands capture much of the relevant
information for photo-z estimation contained in the 30-band data.

Redshift probability density functions
(PDFs) generated by the \emph{Le Phare} template fitting can be used to estimate
redshift uncertainty across the map, letting us identify cells that have
high redshift variance or multiple redshift solutions, as well as
cells with
a 
well-defined mapping to redshift. In Figure~\ref{figure:photz_disp} we show the photo-z dispersion results from the
\emph{Le Phare} code. The dispersion is the modeled
uncertainty in the redshift assigned to each cell, based on the spread
in the cell's redshift
PDF. Figure~\ref{figure:photz_disp} shows that there are well-defined regions in which the modeled
uncertainties are much higher, and that these regions tend to cluster around
sharp boundaries between low- and high-redshift galaxies. Note that
these boundaries are inherent to the data and indicate regions of
significant redshift degeneracy. A
possible improvement in this analysis is to more rigorously estimate
the photometric uncertainty for each cell using a metric for the
volume of color space it represents; we defer this more detailed analysis
to future work.

\subsection{Current spectroscopic coverage in
COSMOS}

One of the most important results of the mapping is that it lets us \emph{directly} test the representativeness of existing spectroscopic
coverage. To do so, we used the master spectroscopic catalog from the COSMOS collaboration (Salvato et al. 2016, in prep).
 The catalog includes redshifts from VLT VIMOS (zCOSMOS,
 \citealp{Lilly07}; VUDS, \citealp{LeFevre15}), Keck MOSFIRE (Scoville et al.
 2015 in prep; MOSDEF, \citealp{Kriek15}), Keck DEIMOS
  (\citealp{Kartaltepe10}, Hasinger et
 al. 2015, in prep), Magellan IMACS \citep{Trump07}, Gemini-S
 \citep{Balogh14}, Subaru FMOS \citep{Silverman15},
 as well as a non-negligible fraction of sources provided by a
 number of smaller programs. It is important
 to note that  the spectroscopic coverage of the COSMOS field is not
 representative of the typical coverage for surveys. Multiple
 instruments with different wavelength coverages and 
 resolutions were employed. Moreover, the spectroscopic programs targeted different types of sources: from AGN to flux-limited
 samples, from group and cluster members to high-redshift
 candidates, etc., providing an exceptional coverage in parameter space.

In the left panel of Figure~\ref{figure:specz}, we show the
map colored by the median spectroscopic redshift of galaxies
associated with each cell, using only galaxies with the highest
confidence redshift assignments (corresponding to $\sim$100\% certainty). The gray regions on the map correspond to cells of color
space for which
no galaxies have such high confidence spectrosopic redshifts; 64\% of
cells fall in this category. In the
right panel of  Figure~\ref{figure:specz} we show the same plot, but
using all confidence $\gtrsim$95\% redshifts in the master
catalog. Significantly more of the galaxy color space is covered with
spectroscopy when the requirement on the quality
of the redshifts is relaxed, with only 51\% of color cells remaining gray. However, for calibration purposes
very high confidence redshifts will be needed, so that the
right-hand panel may be overly optimistic. As can be seen in
both panels,
large and often continuous regions of galaxy color space remain unexplored
with spectroscopy.

\begin{figure}[htb]
        \centering
	\includegraphics[width=0.95\linewidth]{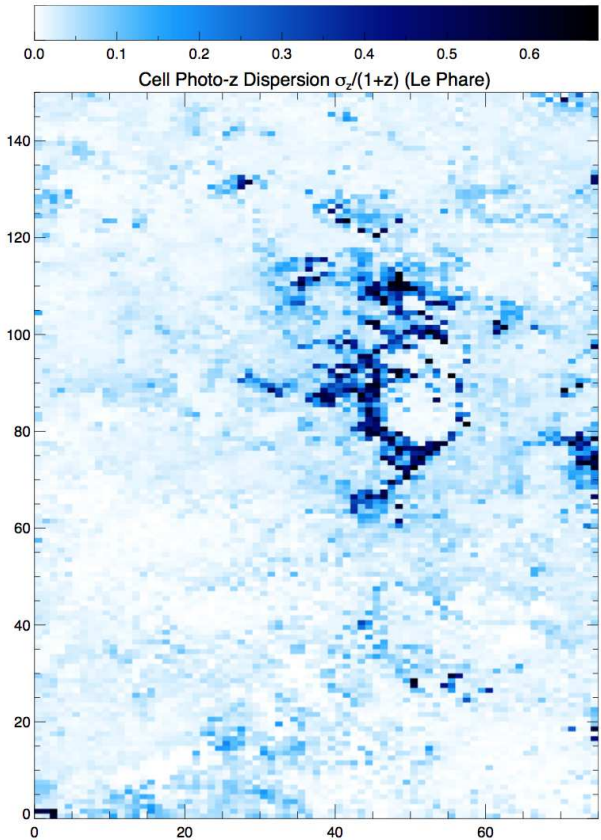}
	\caption{The dispersion in the photo-z computed with the
    \emph{Le Phare} template fitting code as a function of color cell. As can be seen, high
    dispersion regions predominantly fall in localized areas of color
    space near the boundary separating high and low redshift galaxies.}
\label{figure:photz_disp}
\end{figure}

\begin{figure*}[htb]
        \centering
    \includegraphics[width=\linewidth]{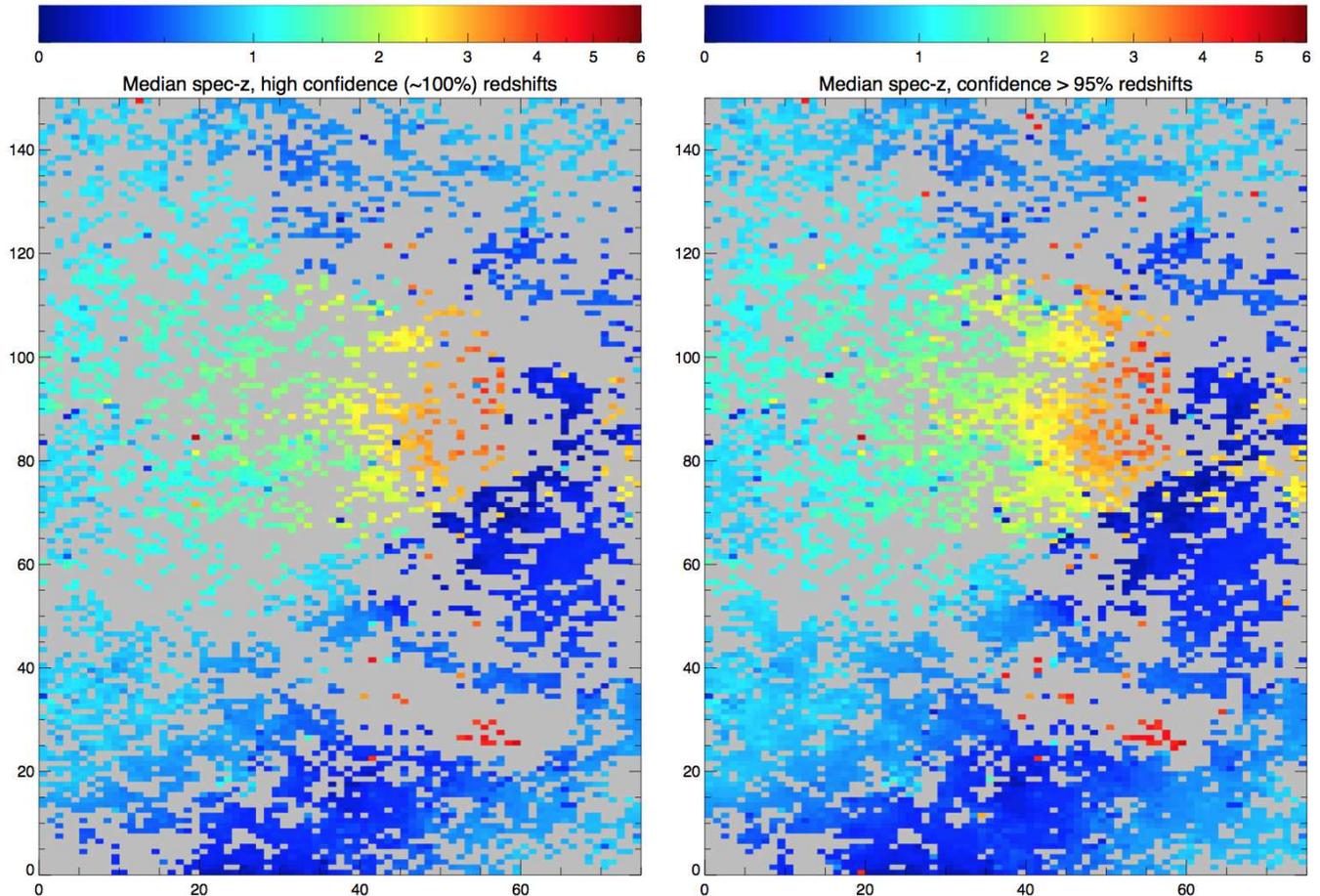} 
	\caption{\emph{Left:} The median spectroscopic
          redshift of galaxies associating with each SOM cell, using
          only very high confidence ($\sim$100\%) redshifts from the COSMOS master spectroscopic
          catalog (Salvato et al., in prep). The redshifts come from a
          variety of surveys that have targeted the COSMOS field; see
          text for details. Gray
          regions correspond to parts of galaxy color
          space for which no high-confidence spectroscopic redshifts
          currently exist. These regions will be of interest for
          training and calibration campaigns. \emph{Right:} The same
          figure, but including all redshifts above $\gtrsim$95\%
          confidence from the COSMOS spectroscopic catalog. Clearly,
          more of the color space is filled in when the quality
          requirement is relaxed, but nevertheless large regions of
          parameter space
         remain unexplored.}
\label{figure:specz}
\end{figure*}

It should be noted that Figure~\ref{figure:specz} is entirely
data-driven, demonstrating the direct association of observed SED with observed redshift. An interesting possibility
suggested by this figure is that the color-redshift relation
may be smoother than expected from photo-z variance estimates from
template fitting (e.g., Figure~\ref{figure:photz_disp}). High
intrinsic variance in the color-redshift mapping should result in large cell-to-cell
variation in median spec-z, whereas the actual distribution appears to
be rather smooth overall.

\begin{figure}[htb]
        \centering
	\includegraphics[width=0.95\linewidth]{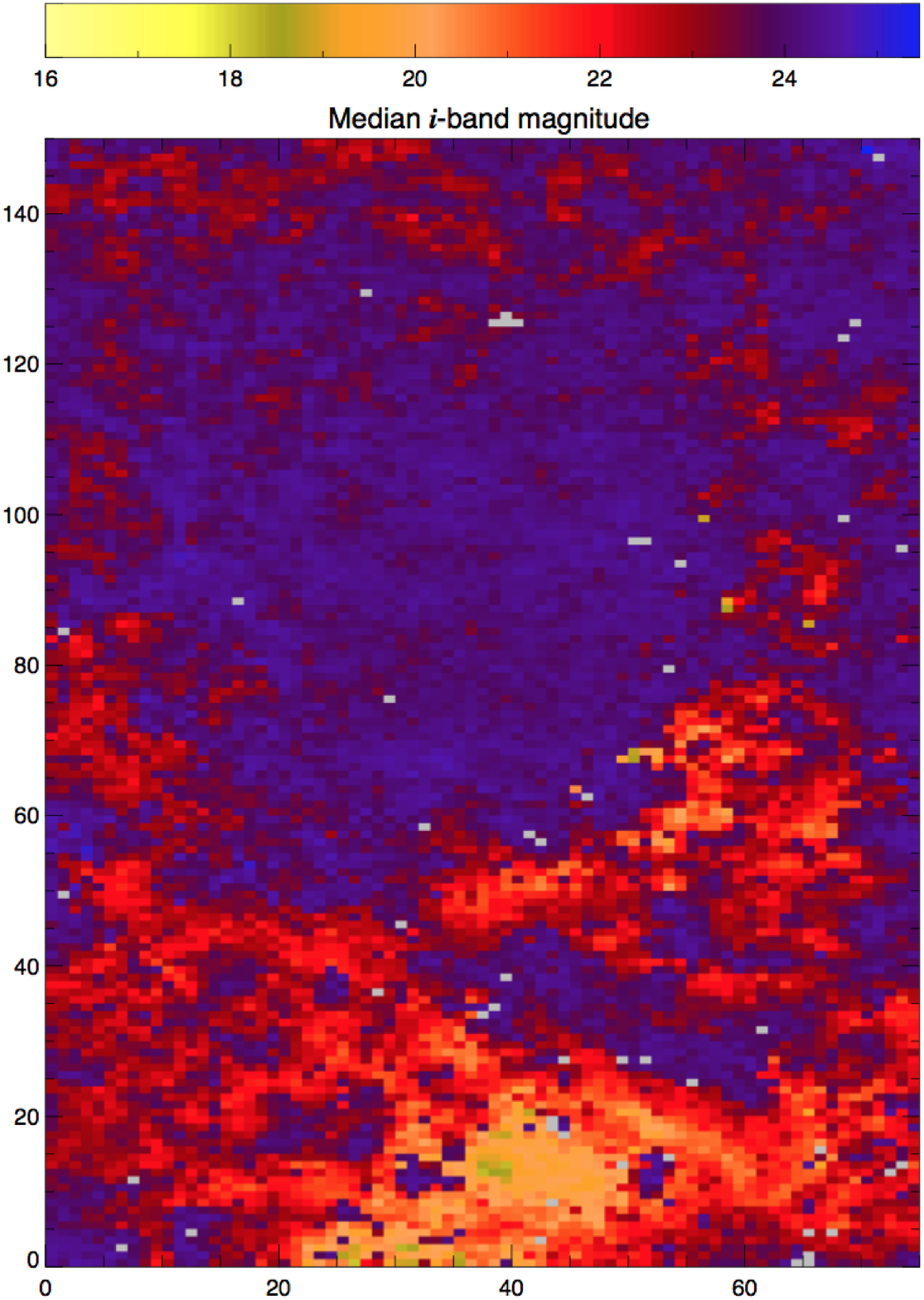}
	\caption{The map colored by the median \emph{i}-band magnitude (AB)
          of galaxies associating with each cell. The strong variation
          of magnitude with color is not unexpected, and largely
          explains the absence of spectra in particular regions of
          galaxy color space.}
\label{figure:magnitude}
\end{figure}

\subsection{Magnitude variation across color space}

Not surprisingly, the median galaxy magnitude varies
strongly with location in color space, as illustrated in 
Figure~\ref{figure:magnitude}. This variation largely determines the
regions of color space that have been explored with spectroscopy, with
intrinsically fainter galaxies less likely to have been
observed. In fact, as we will discuss further in \S6.6, the majority
of galaxies in unexplored regions of color space are faint,
star-forming galaxies at $z\sim0.2-1.5$, which are simply too 
``uninteresting'' (from a galaxy evolution standpoint) to have
been targeted in current spectroscopic
surveys. Such sources will, however, be critically important for 
weak lensing cosmology.

\section{Toward Optimal Spectroscopic Sampling Strategies for Photo-z Calibration}
We have demonstrated that the self-organizing map, when applied to a large photometric dataset, efficiently
characterizes the distribution of galaxies in the
parameter space relevant for photo-z estimation. We now consider the problem of determining the spectroscopic
sample needed to calibrate the $\meanz$ of the tomographic redshift bins
to the required level for weak lensing cosmology. We show that
allocating spectroscopic efforts using the color space mapping can minimize the
spectroscopy needed to reach the requirement on the
calibration of $N(z)$.

\subsection{Estimating the spectroscopic sample needed for calibration}

Obtaining spectroscopic redshifts over the full color space of galaxies is obviously
beneficial, but the question arises: precisely how many spectra are
needed in different regions of color space in order to meet the dark
energy requirement? Here we provide a framework for
understanding this question in terms of the color space mapping.

First we note that each color cell has some subset of galaxies that associate best
with it; let the total number of galaxies associating with the
$i^{\mathrm{th}}$ cell be $n_{i}$. We refer to the true redshift probability
distribution of these galaxies as $P_{i}(z)$. For the sake of this argument we assume that a tomographic
redshift bin for weak lensing will be constructed by selecting all galaxies
associating with some subset of the cells in the SOM. Let the total number
of cells used in that tomographic bin be $c$. Then the true $N(z)$ distribution for galaxies in
the resulting tomographic redshift bin
is: 
\begin{equation} N(z) =
  \sum\limits_{i=1}^cn_{i}P_{i}(z) 
\end{equation} 
The mean of the
$N(z)$ distribution is given by: 
\begin{equation} \langle z \rangle =
  \frac{\int z N(z) dz} { N_{T}} 
\end{equation} 
where the integral is
taken over all redshifts and $N_{T}$ is the total number of galaxies in the
redshift bin. Inserting Equation~(8) into Equation~(9), we find that
the mean redshift of the bin can be expressed as 
\begin{equation}
\begin{split}
  \langle z \rangle = \frac{1}{N_{T}}\int z
  [n_{1}P_{1}(z) + ... + n_{c}P_{c}(z)]dz \\ 
= \frac{1}{N_{T}} [ n_{1}\langle{z_{1}}\rangle + ... +
  n_{c}\langle{z_{c}}\rangle ]
\end{split}
\end{equation}

Equation~(10) is the straightforward result that the mean redshift of the full $N(z)$ distribution is
proportional to the sum of the mean redshifts of each color cell,
weighted by the number of galaxies per cell. The uncertainty in
$\langle z \rangle$ depends on the uncertainty of the mean redshift of
each cell, and is expressed as: 
\begin{equation}
\Delta \langle z \rangle =
\frac{1}{N_{T}}\sqrt{\sum\limits_{i=1}^cn_{i}^2\sigma_{\langle z_{i}
\rangle}^{2} }
\end{equation}

Equation~(11) shows quantitatively what is intuitively clear,
namely that the uncertainty in $\langle z \rangle$ is influenced
more strongly by cells with both high uncertainty in their mean
redshift and a significant number of galaxies associating with them. This indicates
that the largest gain can be realized by sampling more heavily in denser
regions of galaxy color space, as well as those regions with higher
redshift uncertainty. Conversely, cells with very high redshift dispersion
could simply be excluded from the weak lensing sample (although caution would
be needed to ensure that no systematic errors are
introduced by doing so). 

If we assume that the $c$ color cells have roughly equal
numbers of galaxies and that $\sigma_{\langle z_{i}
\rangle}$ is roughly constant across cells, then Equation~(11) becomes:
\begin{equation}
\Delta \langle z \rangle = \sigma_{\langle z_{i}
\rangle}  / \sqrt{c}
\end{equation} With $\sigma_{\langle z_{i} \rangle} \sim 0.05(1+\langle z \rangle)$, we find
$\sim$600 color cells with this level of uncertainty would be
needed to reach the \emph{Euclid} calibration requirement for the redshift
bin. With one spectrum per cell required to reach this level of
uncertainty in $\sigma_{\langle z_{i} \rangle}$,
this
estimate of the number of spectra needed is in rough agreement with that of \citet{Bordoloi10}, and
much lower than estimates for direct calibration through random
sampling. Note that the mean redshifts $\langle z_{i}
\rangle$ for each color cell used in Equation~(10) should
be based on spectroscopic redshifts, to ensure that
the estimates are not systematically biased. The error in a cell's mean
redshift estimate, $\sigma_{\langle z_{i}
\rangle}$, will depend on the dispersion in the $P_{i}(z)$
distribution for the cell, and will scale inversely with the square root of the number of spectra
obtained to estimate it.

The preceding analysis treats the photo-z calibration as a
stratified sampling problem, in which the overall statistics of a
population are inferred through targeted sampling from relatively homogeneous
subpopulations. The gain in statistical precision from using
Equation~(10) to estimate $\langle z \rangle$ can be attributed to the
systematic way in which the full color space is sampled, relative to
blind direct sampling. However, stratified sampling will only outperform
random sampling in the case that the subpopulations being sampled do,
in fact, have lower dispersion than the overall distribution--i.e., in
the case that the $P_{i}(z)$ distributions for the color cells have lower redshift dispersion
than the $N(z)$ distribution of all the galaxies in a
tomographic bin. 

\begin{figure*}[htp]
        \centering
	\includegraphics[width=0.8\linewidth]{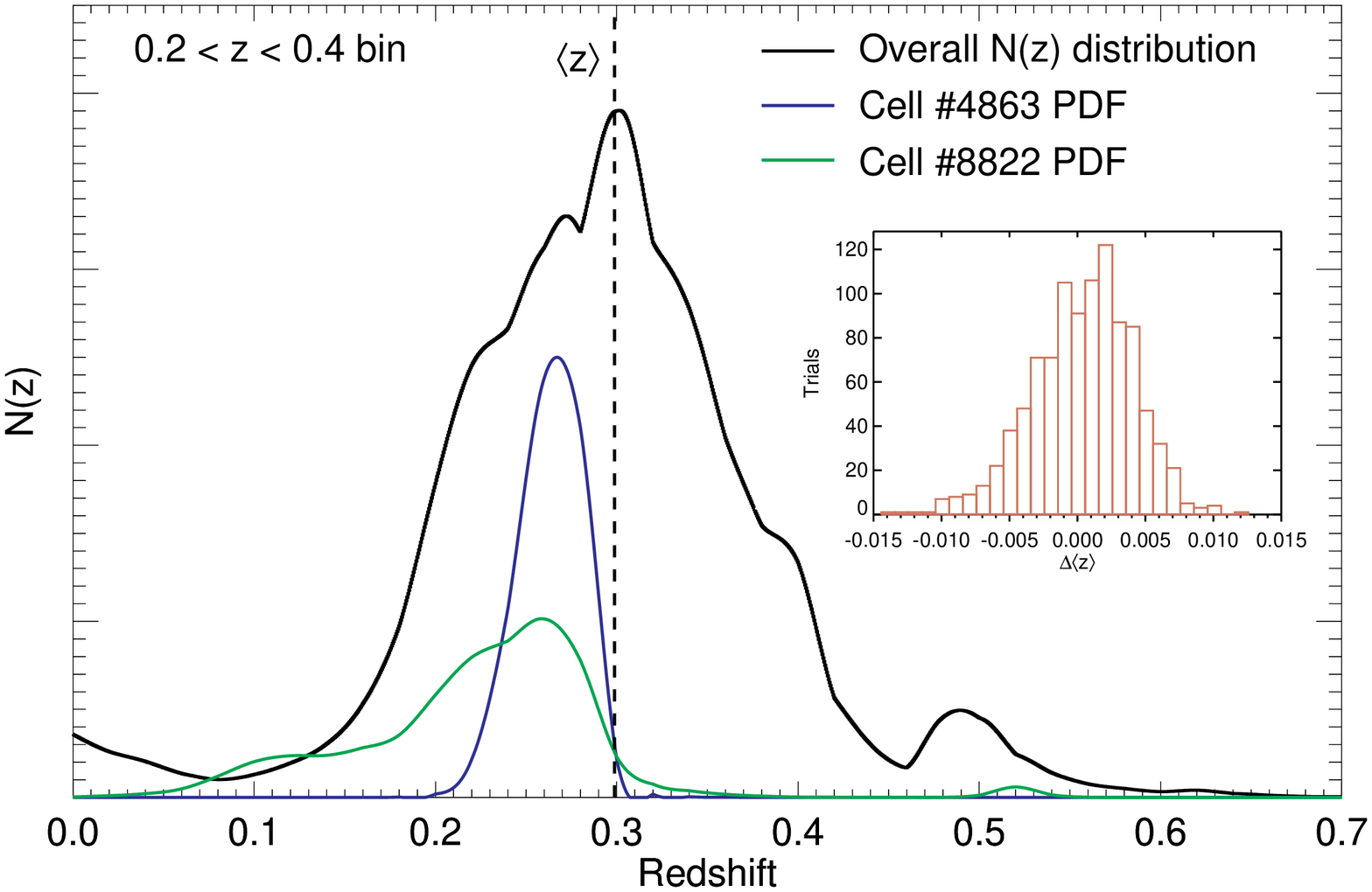}
	\caption{Modeled $N(z)$ distribution for the 0.2-0.4 redshift bin. The
        $N(z)$ distribution is constructed using Equation~(8), treating the $P_{i}(z)$
        functions estimated for each cell from \emph{Le Phare} as truth, and with $n_{i}$
        values from Figure~\ref{figure:occupation}. In addition, two
        random cell PDFs that contributed to the overall $N(z)$ of the tomographic bin are
        shown, one (cell \#4863) with a relatively
        narrowly-peaked distribution and the other (cell \#8822) with more redshift uncertainty. We
        ran Monte Carlo simulations of spectroscopically sampling
        the $N(z)$ distributions
        in various ways to estimate the uncertainty in $\meanz$; see Table~1. The inset plot shows the distribution of
      errors in the estimated $\meanz$ over 1000 Monte Carlo
      trials for the simple strategy of obtaining one spectrum per
      color cell and using Equation~(10) to estimate $\meanz$. The uncertainty in the mean is the standard deviation
      of this distribution, yielding $\sigma_{\meanz}/(1+\meanz)$=0.0028.}
\label{figure:Nz}
\end{figure*}

 \subsection{Simulating different sampling strategies}

Now we attempt to more realistically estimate the spectroscopic coverage needed to
achieve the requirement in our knowledge of $\langle z \rangle$. To
begin, we
assume that the cell redshift PDFs from \emph{Le
  Phare} are reasonably accurate, and can be taken to represent the true
$P_{i}(z)$ distributions for galaxies in each color cell. (This assumption is, of
course, far from certain, and simply serves as a first approximation). With the known
occupation density of cells of the map (Figure~3), we can then use
Equation~(8) to generate realistic $N(z)$ distributions for different
tomographic bins. For this illustration, we break the map up into
photo-z-derived tomographic bins
of width $\Delta z = 0.2$ over $0<z<2$ (although $Euclid$ will most likely use 
somewhat different bins in practice). An example of one of the $N(z)$
distributions modeled in this way is
shown in Figure~\ref{figure:Nz}.

The uncertainty in the estimated $\meanz$ of these $N(z)$
distributions can then be tested for different spectroscopic
sampling strategies through Monte Carlo simulations, in which 
spectroscopy is simulated by randomly drawing from the $P_{i}(z)$ 
distributions. (Alternatively, given
our knowledge of the individual $\sigma_{\langle z_{i}
\rangle}$ uncertainties, Equation~(11) can be used directly. In fact, the results were checked in both
ways and found to be in agreement).

The results of three possible sampling strategies are given in
Table~1. The simplest strategy tested (``Strategy 1'') is to obtain one spectrum per color cell
in order to estimate the cell mean redshifts. Equation~(10) is then used to compute the overall mean of the
tomographic bin. We expect to meet the $Euclid$ requirement, $\Delta
\langle z \rangle \leq 0.002(1+\langle z \rangle)$, for 3/10
bins (and come close in the others) with this approach, which
would require $\sim$11k spectra in total.

The second strategy tested is similar to the first, in that one
spectrum per cell is obtained. However, galaxies associated with the
5\% of the cells in each bin with the
highest redshift uncertainty are rejected from the weak lensing sample,
and these cells are ignored in the sampling. This
significantly reduces the uncertainty in the $\meanz$ estimates, with 6/10 bins
meeting the requirement; moreover, it reduces the total number of spectra
needed by 5\%. However, it comes at the cost of reducing the number of
galaxies in the weak lensing sample.

The third strategy is to sample the 5\% of the cells with the highest
redshift uncertainty with three spectra each in order to estimate
their mean redshifts with greater accuracy, again obtaining one spectrum for the other
95\% of the cells. This strategy again lowers the uncertainty in the
$\meanz$ estimates substantially, but at the cost of increased
spectroscopic effort, requiring $\sim$12k spectra in total. The
additional spectra needed may also prove to be the more difficult ones
to obtain, so the effort needed cannot be assumed to scale linearly
with the number of spectra.

These examples are simply meant to be illustrative of the possible strategies
that can be adopted for the spectroscopic calibration. More refined
strategies are possible -- for
example, an optimal allocation of spectroscopic effort could be
devised that scales the number of spectra in a given region of color space
proportionately to the redshift uncertainty in that region, while rejecting limited regions of color space that are both
highly uncertain and difficult for spectroscopy. Additional
spectroscopy may need to be allocated to the higher redshift bins, for
which there tend to be fewer cells overall as well as higher
dispersion within cells. Tomographic bins could also be intentionally generated to
minimize the uncertainty in $\meanz$. The simpler examples shown here do illustrate that, if we believe the cell
$P_{i}(z)$ estimates from template fitting, the $Euclid$ calibration requirement $\Delta
\langle z \rangle \leq 0.002(1+\langle z \rangle)$ is achievable with
$\sim$10-15k spectra in total (roughly half of which already
exist).

\subsubsection{Is filling the map with spectroscopy necessary?}
The number of spectra needed derived above assumes that at least one
spectrum per SOM color cell is necessary to estimate the
$\langle z_{i} \rangle$ for that cell. However, if a particular region of color space is very well
understood and maps smoothly to redshift, sparser spectroscopic
sampling in that region together with interpolation across cells might be sufficient. Equivalently, groups of neighboring cells with low redshift
uncertainty that map to roughly the same redshift could potentially be merged
using a secondary clustering procedure, thus lowering the overall number of
cells and the number of spectra required. These considerations
 suggest that, while the exact number of spectra required to meet the calibration
requirement is uncertain, the results presented above are likely
to represent upper limits.


\begin{center}
\begin{deluxetable*}{c|cc|ccc|cc}
\tabletypesize{\scriptsize}
\tablewidth{0pt} 
\tablecolumns{8}
\tablecaption{Simulated uncertainty in $\meanz$ for representative
  redshift bins for different sampling strategies.}
\tablehead{
 \colhead{} &
  \multicolumn{2}{c}{Strategy 1\tablenotemark{a}} &
  \multicolumn{3}{c}{Strategy 2\tablenotemark{b}} &
  \multicolumn{2}{c}{Strategy 3\tablenotemark{c}} \\ [0.1cm] \hline \noalign{\vskip 0.1cm}
 \colhead{Redshift bin} &
  \colhead{\#Spectra} &
  \colhead{$\sigma_{\meanz}/(1+\meanz)$} &
  \colhead{\#Spectra} &
  \colhead{\% Sample lost\tablenotemark{d}} &
  \colhead{$\sigma_{\meanz}/(1+\meanz)$} &
  \colhead{\#Spectra} &
  \colhead{$\sigma_{\meanz}/(1+\meanz)$} 
}
\startdata 
 0.0-0.2 & 659 & 0.0034 & 627 & 4.2 & 0.0024 & 723 & 0.0028  \\
 0.2-0.4 & 1383 & 0.0028 & 1314 & 4.6 & 0.0015 & 1521 & 0.0020 \\
  0.4-0.6 & 2226 & 0.0014 & 2115 & 3.9 & 0.0007 & 2448 & 0.0010 \\
  0.6-0.8 & 2027 & 0.0018 & 1926 & 4.3 & 0.0005 & 2229 & 0.0012 \\
  0.8-1.0 & 1357 & 0.0021 & 1290 & 4.4 & 0.0009 & 1491 & 0.0013 \\
  1.0-1.2 & 1705 & 0.0011 & 1620 & 4.6 & 0.0005 & 1875 & 0.0008  \\
  1.2-1.4 & 559 & 0.0029 & 532 & 4.4 & 0.0015 & 613 & 0.0021 \\
  1.4-1.6 & 391 & 0.0044 & 372 & 3.3 & 0.0021 & 429 & 0.0031 \\
  1.6-1.8 & 268 & 0.0064 & 255 & 2.7 & 0.0050 & 294 & 0.0055 \\
  1.8-2.0 & 164 & 0.0093 & 156 & 2.1 & 0.0085 & 180 & 0.0088 \\
  [0.1cm] \hline \noalign{\vskip 0.1cm}
Total \#spectra: & \multicolumn{2}{c}{10739} &
\multicolumn{3}{c}{10207} & \multicolumn{2}{c}{11793}
\enddata
\tablenotetext{a}{Obtaining one spectrum per color
  cell to estimate $\langle z_{i} \rangle$, with $\meanz$ computed using Equation (10).}
\tablenotetext{b}{Again obtaining one spectrum per color
  cell to estimate $\langle z_{i} \rangle$, but rejecting the
  5\% of cells with the highest redshift uncertainty.}
\tablenotetext{c}{Obtaining three spectra per color
  cell for the
  5\% of cells with the highest redshift uncertainty, one spectrum per
cell for the other 95\%.}
\tablenotetext{d}{The fraction of galaxies lost from the weak lensing
  sample for that tomographic bin due to excluding 5\% of the most
  uncertain color cells.}
\label{table:sigz}
\end{deluxetable*}
\end{center}

\subsection{Estimating the true uncertainty in the color-redshift mapping}

\begin{figure*}[htb]
        \centering
	\includegraphics[width=0.95\linewidth]{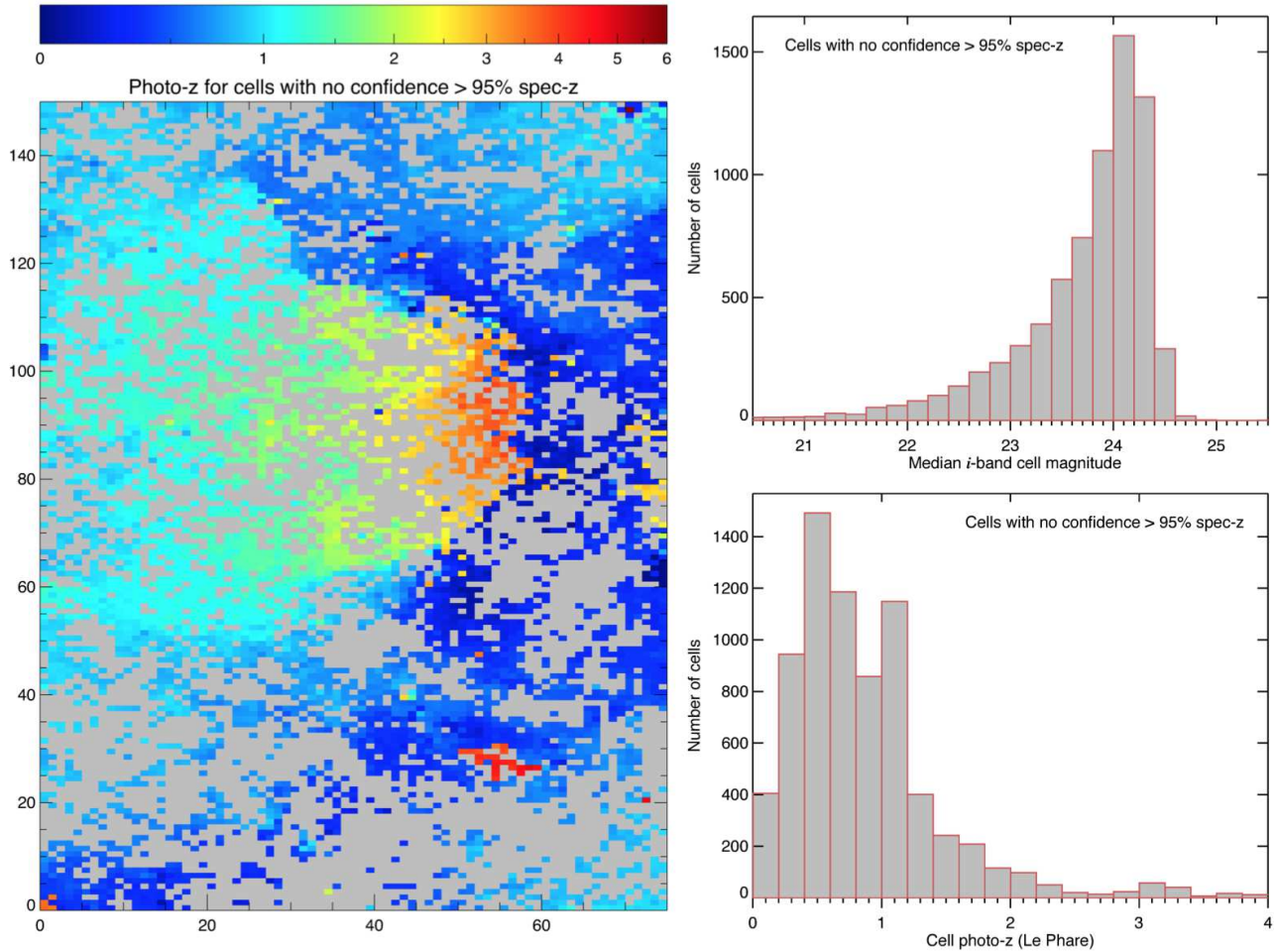}
	\caption{\emph{Left:} The inverse of the right panel of
          Figure~\ref{figure:specz}, illustrating the distribution and
          photometric redshifts of color cells currently containing no
          galaxies with confidence $>$95\% redshifts. \emph{Right, top:} Magnitude distribution
    of cells unsampled by
    spectroscopy, where the cell magnitude is defined as the median
    \emph{i}-band magnitude (AB) of galaxies associating with the 
    cell. \emph{Right, bottom:} Photo-z distribution of unsampled cells,
    computed with \emph{Le Phare} on the 8-band data representative of
  the \emph{Euclid} photometry. The majority of the
    color space regions currently unsampled by spectroscopy correspond
    to faint galaxies ($i~\mathrm{band}\sim~23-24.5$~AB) at
    $\mathrm{z}\sim0.2-1.5$.}
\label{figure:no_specz}
\end{figure*}

The analysis above highlights the important role played by the
true uncertainty in the mapping from color to redshift for
some number of broadband filters. A single spectroscopic redshift gives us an estimate of a cell's mean redshift with an
uncertainty that depends on the true dispersion in $P_{i}(z)$ for the
cell. Unfortunately, we cannot know this distribution precisely without heavily
sampling the cell with spectroscopy, which is impractical (we can,
however, model it with different photo-z codes). 

Given the importance of the uncertainty in the mapping of color
to redshift in different parts of color space, strategies to constrain
this uncertainty efficiently should be considered. One possibility is that a limited amount of ancillary photometry
can effectively identify the redshift variation within cells. The
reason this could work is that objects
with very different redshifts but
similar \emph{Euclid} colors are likely to be 
distinguishable in other bands (e.g., IR or FUV). Moreover,
well-defined and distinct magnitude distributions for objects in the
same region of color space could indicate and help break
a color-redshift degeneracy.


Another interesting possibility is that the uncertainty in $P_{i}(z)$ in
different parts of color space can be constrained
\emph{from the map itself}, as it is filled in with spectroscopy. This
is because the cell-to-cell redshifts would be
expected to show high
variation in parts of color space where the relation has high
intrinsic variation, and vary more smoothly in regions where the
relation is well-defined. We defer a detailed analysis
of this possibility to future work. 

\subsection{Effect of photometric error on localization in color space}

Photo-z uncertainty is due both to the
inherent uncertainty in the mapping from some number of broadband colors to redshift, as well
as the uncertainty in the colors themselves due to photometric
error. It is well-known that photometric redshift performance degrades
rapidly at low signal-to-noise for the latter reason. 

\emph{Euclid} and other dark energy surveys will also observe deep calibration fields, in
which the survey depth is $\sim$2 magnitudes deeper than the main
survey. These will preferentially be the fields with
spectroscopic redshifts used for training and calibration. Because of
the photometric depth, the photometric error will be
negligible in these fields, and the uncertainty in mapping
color to redshift will be due to inherent uncertainty in the
relation. 

Even if the relation between color and redshift
is mapped as fully as possible in the deep fields, photometric error in the shallower
full survey will introduce
uncertainties by allowing galaxies
to scatter from one part of color space to another. The errors thus
introduced to the tomographic redshift bins can be well characterized using the multiple observations of the deep fields, and folded into
the estimates of $\sigma_{\langle z_{i} \rangle}$. The ultimate effect
on the $N(z)$ estimates will depend on the S/N cut used for the weak
lensing sample.

\subsection{Cosmic variance}

One of the primary difficulties with direct measurement of the $N(z)$ distribution
for tomographic redshift bins is the need for multiple independent sightlines in order to avoid cosmic
variance-induced bias in the $N(z)$ estimates. Systematically measuring the color-redshift
relation as described here, however, largely sidesteps the problem posed by cosmic variance. This is because the true $\rho{(\vec{C})}$
distribution can be inferred from the full survey (which will be
unaffected by cosmic variance or shot noise), while the
calibration of $P(z|\vec{C})$ can be performed on data from a small number of
fields, as long as galaxies in those fields span the overall galaxy color
space sufficiently. 

\subsection{Galaxies in under-sampled regions of color space}

From the preceding analysis, a reasonable step toward calibration of the photo-z's
for cosmology is to target the regions of multicolor space
currently lacking spectroscopy (the gray regions in
Figure~\ref{figure:specz}). It is therefore important to understand
the nature of the galaxies in these regions, in order to predict the spectroscopic effort needed. 


Of the 11,250 cells in the SOM presented here, roughly half currently have no
objects with high-confidence spectroscopic redshifts. The distribution
of these cells on the map, as well as their photometric redshift
estimates, are displayed on the left side of
Figure~\ref{figure:no_specz}. The right side of Figure~\ref{figure:no_specz} shows the overall magnitude
and photometric redshift distribution of the unsampled cells of color space. Most
unsampled cells represent galaxies fainter than $i=23$~(AB) at redshifts
$\mathrm{z}\sim0.2-1.5$, and $\sim$83\% of these are classified as
star-forming by template fitting. These magnitude, redshift,
  and galaxy type estimates directly inform our prediction
  of the spectroscopic effort that will be required to
  calibrate the unsampled regions of galaxy color space.

Generally speaking, these galaxies have not been
targeted in existing spectroscopic surveys because they are
faint and not considered critical for galaxy evolution
studies. However, they are abundant and thus important for weak lensing
cosmology. In Appendix A we give a detailed estimate of the
observing time that would be needed to fill in the empty parts of color
space with a fiducial survey with Keck, making use of LRIS, DEIMOS
and MOSFIRE. We find that $\sim$40 nights would
be required if we reject the 1\% most difficult cells -- a large time allocation, but not unprecedented in
comparison with other large spectroscopic surveys. This is significantly less than the 
$\sim$100 nights needed to obtain a truly representative sample without prior knowledge 
of the color distribution \citep{Newman15}.  For both LSST and
$WFIRST$ the calibration sample required is likely to be significantly
larger, due to the greater photometric depths of these surveys in
comparison with $Euclid$. Therefore, methods to improve the sampling as proposed here 
will be even more important to make the problem tractable for those surveys.  
 

\section{Discussion}

Statistically well-understood photometric redshift estimates for billions of galaxies will
be critical to the success of upcoming Stage IV dark energy surveys.  We have demonstrated that
self-organized mapping of the multidimensional color distribution of
galaxies in a broadband survey such as \emph{Euclid} has significant benefits
for redshift calibration. Importantly, this
technique lets us identify regions of the 
photometric parameter space in which the density of galaxies $\rho{(\vec{C})}$ is
non-negligible, but spectroscopic redshifts do not
currently exist. These unexplored
regions will be of primary interest for spectroscopic training and
calibration efforts. 

Applying our SOM-based analysis to the COSMOS field, we show that the
regions of galaxy parameter space currently lacking spectroscopic coverage
generally correspond to faint (\emph{i}-band magnitude (AB) $\gtrsim$ 23), star-forming
galaxies at $z<2$. We estimated the spectroscopy required to fill the color
space map with one spectrum per cell (which would come close to or achieve the required precision for
calibration) and found that a targeted, $\sim$40~night campaign with Keck
(making use of LRIS, DEIMOS and MOSFIRE)
would be sufficient (Appendix A). It should be noted that this analysis is specific
to the $Euclid$ survey. The calibration needs of both LSST and
\emph{WFIRST} are likely to be greater, due to the deeper photometry
that will be obtained by those surveys. 

We demonstrated that systematically sampling the color space occupied by
galaxies with spectroscopy can efficiently constrain the $N(z)$ distribution of galaxies in
tomographic bins. The precise number of spectra needed to meet the bias requirement in $\meanz$
for cosmology depends
sensitively on the uncertainty in the color-redshift mapping. Template-based estimates suggest that this uncertainty is rather high in
some regions of \emph{Euclid}-like color space. However, the smoothness of the
spectroscopic redshift distribution on the map suggests that the
template-based uncertainties may be overestimated, which would reduce the total number of spectra
needed for calibration. 

Assuming that the uncertainties in $P(z|\vec{C)}$ from
template fitting are accurate, we demonstrate that the $Euclid$
requirement on $\Delta \langle z \rangle$ should be achievable
with $\sim$10-15k total spectra, about half of which already exist
from various spectroscopic surveys that have targeted the COSMOS
field. Understanding the true uncertainty in $P(z|\vec{C})$ will likely prove critical to constraining the
uncertainty in $\langle z \rangle$ for the tomographic bins, and we
suggest that developing efficient ways of constraining this
uncertainty should be prioritized.

The topological nature of the self-organizing
map technique suggests other possible uses. 
For example, a potentially very useful aspect of the SOM is that it lets us
quantify the ``normality'' of an object by how
well-represented it is by some cell in the map. Rare
objects, such as AGN, blended sources, or objects with otherwise contaminated photometry could
possibly be identified in this way. We also
note that the mapping, by empirically constraining the galaxy
colors that appear in the data, can be used both to generate
consistent priors for template fitting codes as well as test the
representativeness of galaxy template sets. These applications will be
explored in future work.

\acknowledgements

We thank the anonymous referee for constructive comments that significantly improved this
work. We thank Dr. Ranga Ram Chary, Dr. Ciro Donalek, and Dr. Mattias Carrasco-Kind for useful
discussions. D.M., P.C., D.S., and J.R., acknowledge support by NASA
ROSES grant
12-EUCLID12- 0004. J.R. is supported by JPL, run by Caltech for NASA. H.Ho. is supported by the DFG Emmy Noether grant Hi
1495/2-1. S.S. was supported by Department of Energy Grant DESC0009999. Data from the VUDS survey based on data obtained with the European Southern Observatory
Very Large Telescope, Paranal, Chile, under Large Program
185.A-0791. This work is based in part on data products made available at the CESAM
data center, Laboratoire d’Astrophysique de Marseille.

\bibliographystyle{apj}
\bibliography{biblio}

\appendix

\section{A. Estimating the observing time required for the \emph{Euclid} calibration}

Given the $<$0.2\% accuracy in $\meanz$ required for the \emph{Euclid} 
tomographic bins, and following the
analysis presented above, a nearly optimal approach would be to
obtain one spectrum per SOM cell, while rejecting $\sim$1\% of the cells requiring the longest
spectroscopic observations. Taking existing spectroscopy into 
account, a total of $\sim$5k new spectroscopic redshifts would be
needed. We
estimate that these spectra could be obtained in $\sim$40 nights with Keck, as outlined
below.

To quantify the required exposure time, we constructed a fiducial survey on the Keck
telescope with the Low Resolution Imaging Spectrograph (LRIS) \citep{Oke95}, the Deep
Extragalactic Imaging Multi-Object Spectrograph (DEIMOS) \citep{Faber03}, and the Multi-Object
Spectrograph for Infrared Exploration (MOSFIRE) \citep{McLean12} instruments.  This
telescope/instrument combination was chosen because the full redshift
range of the calibration sample can be
optimally probed with these instruments, and their performance in
obtaining redshifts for \emph{i}$\sim$24.5 galaxies has been demonstrated in
numerous publications (e.g., \citealp{Steidel04, Newman13, Kriek15}).  For LRIS we follow
\citet{Steidel04} and
assume the 300 groove mm$^{-1}$ grism blazed at 5000\AA\ on the blue side and the 600
groove mm$^{-1}$ grating blazed at 10,000\AA\ on the red side with the D560
dichroic.  With DEIMOS the 600 grove mm$^{-1}$ grating tilted to 7000\AA\ was
assumed.  MOSFIRE was assumed to be in its default configuration.  Sensitivities were
estimated using the official exposure time calculators (ETCs) provided by Keck by
scaling from a 24th magnitude flat spectrum object.  We assume
1$\arcsec$ seeing, a 1$\arcsec$ wide slit, an airmass of 1.3, and we
include appropriate slit losses.  For all instruments we scaled the SNR to a binning of \emph{R}$\sim$1500, the
minimum required resolution for calibration redshifts.  The assumed
SNRs in a one hour exposure
at 24th magnitude (AB)
are given in Table~\ref{table:sens}.

We assume that the galaxies in the cells needing
spectroscopy have the redshifts, galaxy spectral
types, and reddenings derived from template fitting with \emph{Le Phare}. The modeled
galaxy spectral types, redshifts, and observed magnitudes were used to determine the required SNR and the
instrument such that a $>$99\% reliable redshift can be obtained.  For star
forming galaxies at $z<2.7$ we require SNR~=~2 on the
continuum because bright rest-frame optical emission lines will be used to determine
the redshift. For star forming galaxies at $z>=2.7$ we require
SNR~=~3 on the continuum to clearly detect the Lyman break and the rest
frame ultraviolet (UV) absorption features with LRIS or DEIMOS (e.g.,
\citealp{Steidel03}).  For galaxies classified as passive, we require SNR~=~5 on the
continuum (e.g., \citealp{Kriek09, Onodera12}), while objects intermediate between passive and star-forming were allowed to
linearly scale between an SNR of 5 and 2 with increasingly
star-forming spectral template, because the spectral feature strength increases with star formation rate.

The magnitude measured in the band closest to the most prominent spectral feature
was assumed for the SNR calculation, and the instrument with the highest sensitivity
at that feature was assumed.  For passive galaxies it was assumed that the 4000\AA\
break must be targeted at $z<2.3$ and the 1216\AA\ Lyman forest break at higher
redshifts, with DEIMOS used at $z<1.3$, LRIS at $1.3<z<1.4$, MOSFIRE at $1.4<z<2.3$,
LRIS at $2.3<z<3.5$ and DEIMOS at $z>3.5$.  For other galaxies, the strongest of
H$\alpha$, H$\beta$, O[III] and O[II] was targeted at $z<2.7$, with
DEIMOS at $z<1.5$ and MOSFIRE at
$1.5<z<2.7$. The 1216\AA\
Lyman forest break was targeted at higher redshifts, with LRIS at $2.7<z<3.5$ and DEIMOS at $z>3.5$. 

Objects were
then grouped into masks by instrument and exposure time, assuming a
multiplexing of 70 for DEIMOS and 20 for LRIS and MOSFIRE, making the assumption
that deep observations could be obtained for rare faint objects by observing them in
multiple masks.  Assuming nights are 10h long, overheads are 10\%, 20\% of the
objects need to be observed by more than one instrument to confirm the redshift,
and 30\% losses due to weather, we obtain the estimate of required observing time
given in Table~\ref{table:sens}.  

An exploratory program in early 2015 used samples from poorly sampled regions of
color space as fillers on 2-4 hr Keck DEIMOS slit masks, finding that $>$98\% of sources were readily identified from
strong [OII], [OIII], and/or H$\alpha$ emission, while the non-detected sources had photometric redshifts
for which no line detection was expected by DEIMOS.

We note that an additional $\sim$12 nights would be required to get to
99.8\% completeness in color cells, and $\sim$49 (for a total of
$\sim$100) more nights to reach 99.9\% completeness.  This confirms
the difficulty in obtaining truly complete samples noted by previous
work, as well as the importance of systematically rejecting sources \citep{Newman15}.

\begin{center}
\begin{deluxetable*}{lccc}
\tablewidth{0.8\textwidth} 
\tablecolumns{4}
\tablecaption{Here we give the assumed continuum sensitivity per \emph{R}$\sim$1500 resolution element for selected Keck
  instruments on a flat-spectrum 24th magnitude (AB) object in a
  one-hour exposure.  In the last column we give the estimated
  number of nights required for each
 instrument in a fiducial survey designed to complete the \emph{Euclid}
color space calibration.}
\tablehead{
 \colhead{Instrument} &
 \colhead{Band} &
  \colhead{SNR} &
\colhead{Number of nights}}
\startdata 
LRIS & I& 1.5 & 7\\
DEIMOS & I & 2.0 & 19\\
MOSFIRE & Y & 0.7 & 4\\
MOSFIRE & J & 0.6 & 1\\
MOSFIRE & H & 0.5 & 7\\
MOSFIRE & K & 0.4 & 1\\
\enddata
\label{table:sens}
\end{deluxetable*}
\end{center}

\section{B. Alternate SOM Examples}
\begin{figure*}[htb]
        \centering
        \includegraphics[width=\linewidth]{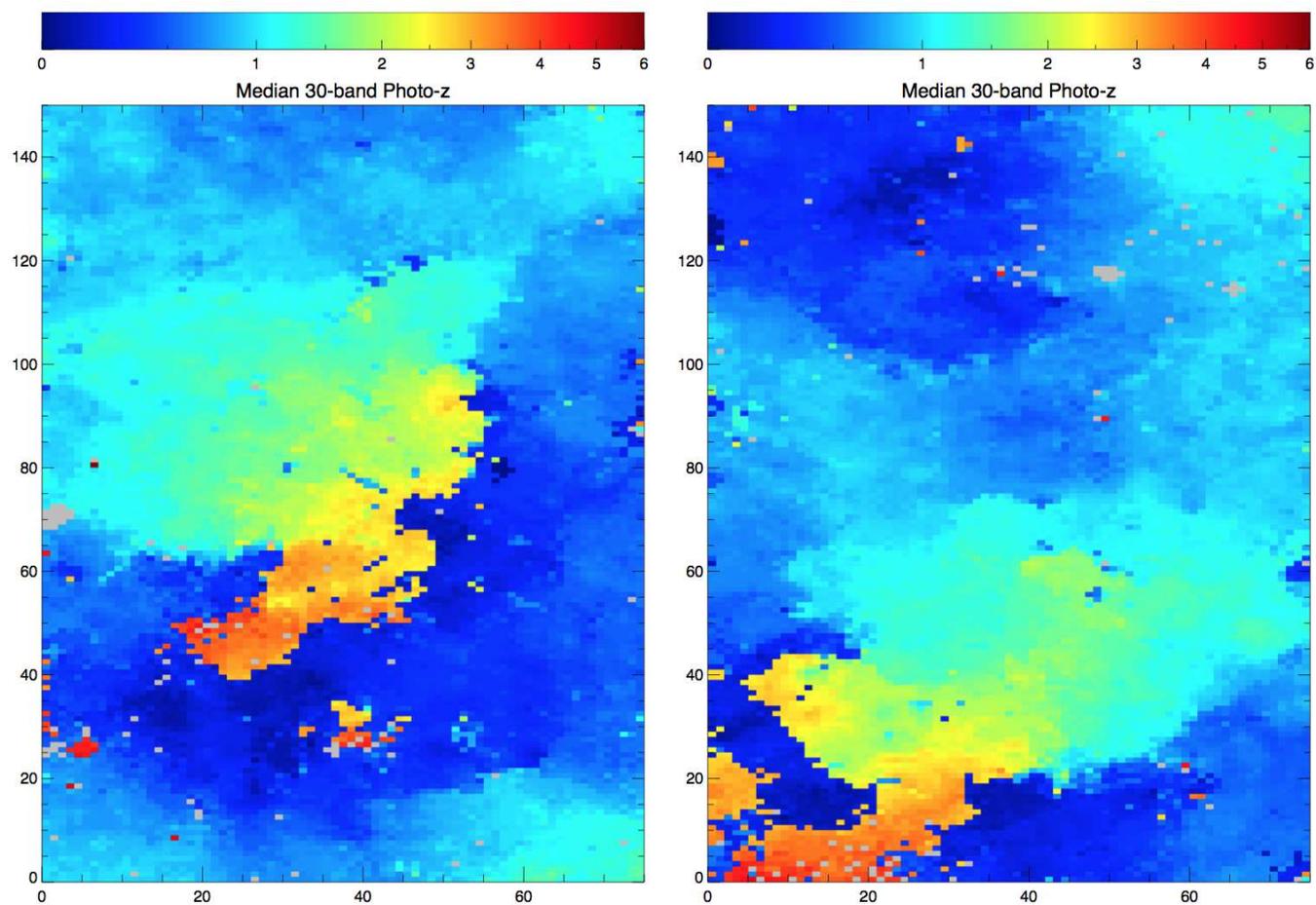} 
	\caption{Two alternate versions of the SOM created with
          different initial conditions/training order, colored
          by the cell 30-band median photo-z (compare to Figure~4, right panel).}
\label{figure:alt_som}
\end{figure*}

Figure~\ref{figure:alt_som} shows two alternate maps generated with
the same COSMOS data, but with different starting conditions and training orders. Note that the overall
topological features are the same. The representativeness of these
maps (in the sense described in \S4.3) are essentially identical to
each other and the map shown throughout the paper. However, the positions and orientations of different
         photometric clusters are random.

\end{document}